# Securing the world's software

By incorporating automation into workflows, communities are making it easier and faster to identify and remediate vulnerabilities.

GitHub

The 2020 State of the
# OCTOVERSE

03

# Securing the world's software

In this report, we investigate open source security: how many projects rely on open source software, the likelihood of having a vulnerability, and best practices for remediation.

## //table of contents

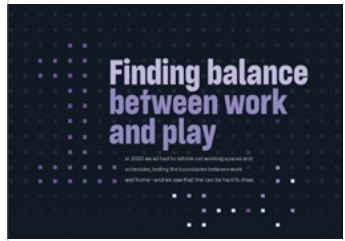

Finding Balance
**Productivity report** →

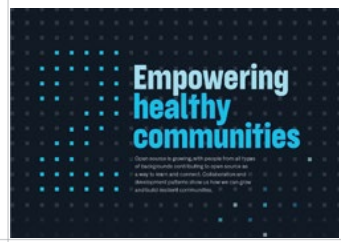

Empowering communities
**Community report** →

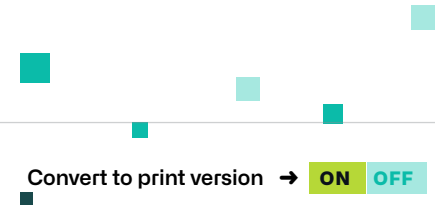

Convert to print version → ON OFF

# Executive summary

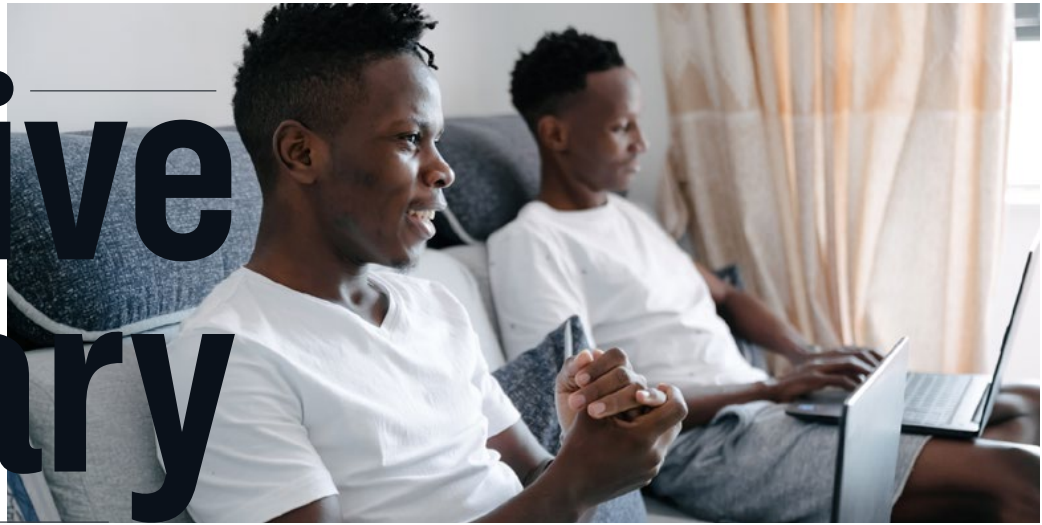

**59%**

chance of getting a security alert in the next year on any active repository with supported package ecosystems

Open source is the connective tissue for much of the information economy. You would be hard-pressed to find a scenario where your data does not pass through at least one open source component. Many of the services and technology we all rely on, from banking to healthcare, also rely on open source software. The artifacts of open source code serve as critical infrastructure for much of the global economy, making the security of open source software mission-critical to the world.

//executive summary

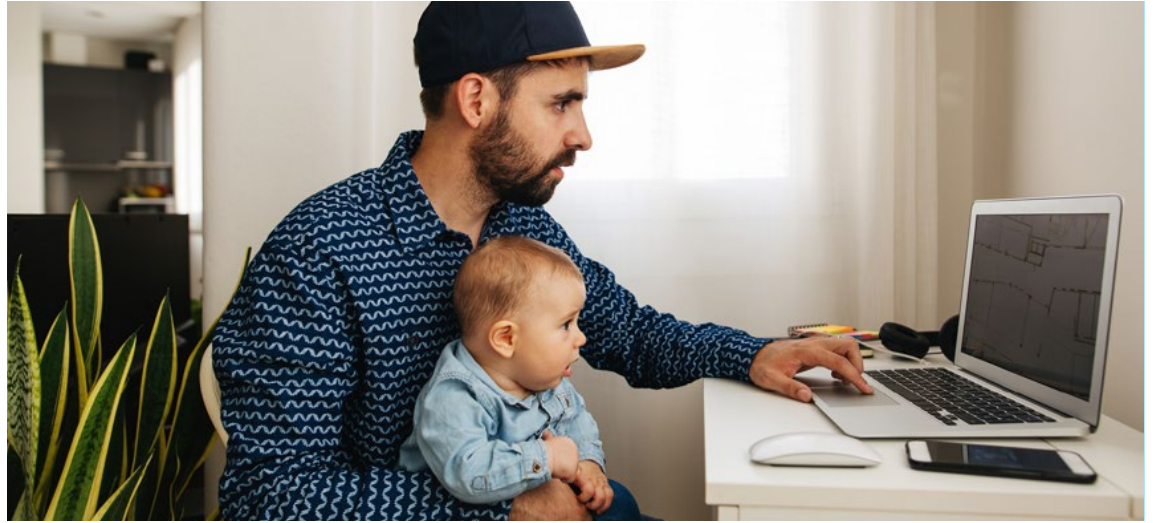

As the largest platform for open source in the world, GitHub is in a unique position to analyze open source software dependencies and the impact of vulnerabilities in those dependencies, and to alert users to address them at scale. Our visibility into vulnerability reporting, alerting, and remediation at GitHub scale allows us to identify important trends in open source security.

The analysis in this section of the Octoverse report pulls together a unique and cohesive picture of open source security and the lifecycle of a vulnerability, identifying key opportunities where we, as a community, can improve the security of open source. It also identifies areas where software teams can focus resources to improve.

# 17%

of vulnerabilities are explicitly malicious but triggered just 0.2% of alerts

//executive summary

# Key findings

**01**

### Most projects on GitHub rely on open source software.
We see the most frequent use of open source dependencies in JavaScript (94%), Ruby (90%), and .NET (90%).

**02**

### Active repositories with a supported package ecosystem have a 59% chance of getting a security alert in the next 12 months.
Ruby (81%) and JavaScript (73%) repositories were the most-likely to receive an alert in the last 12 months. Our analysis also breaks down advisories by severity.

**03**

### Security vulnerabilities often go undetected for more than four years before being disclosed.
Once they are identified, the package maintainer and security community typically create and release a fix in just over four weeks. This highlights the opportunities to improve vulnerability detection in the security community.

**04**

### Most software vulnerabilities are mistakes, not malicious attacks.
Analysis on a random sample of 521 advisories from across our six ecosystems found that 17% of the advisories were related to explicitly malicious behavior such as backdoor attempts. These malicious vulnerabilities were generally in seldom-used packages, but triggered just 0.2% of alerts. While malicious attacks are more likely to get attention in security circles, most vulnerabilities are caused by mistakes.

**05**

### Automation accelerates open source supply chain security.
Repositories that automatically generate a Dependabot pull request patch their software 13 days sooner, or 1.4 times faster, than those that don't. This is one way that teams can "shift left," by building security into development workflows and amplifying the impact of security findings.

# Take action

Take these actions to protect against vulnerabilties.

**01**

### Check your dependencies for vulnerabilities regularly.
The first step is knowing, and you can't patch what you don't know about. Few people have alerts enabled for private repositories, but that can leave you open to threats. With automated alerting, companies and open source projects can stay up to date on security vulnerabilities, information, and patches.

**02**

### Participate in the community if you have a security team.
Open source is critical infrastructure, and we should all contribute to the security of open source software. One way to contribute is by looking for security vulnerabilities in the open source code you use, and reporting any you find privately to the maintainers. Another way to contribute is by using CodeQL to search your own code for vulnerabilities, then share your query to help others do the same.

**03**

### Use automation to remediate vulnerabilities and stay secure.
Using automated alerting and patching tools to secure software quickly means attack surfaces are evolving, making it harder for attackers to exploit. Repositories that automatically generate pull requests to update vulnerable dependencies patch their software 1.4 times faster than those who don't. Automating security practices helps your team secure your code as developers share their expertise with their community, remove security and engineering silos, and scale their expertise.

**04**

### Remediate vulnerabilities quickly and keep your code base current.
Patch your software early and often to secure it with known security remediations. Delays in remediation can leave you open to exploits, and may cause difficulty with future patches that rely on previous updates. You should also update your codebase to the latest version in a timely manner to benefit from security updates and community expertise. Small delays quickly add up to years, and falling behind can make a significant difference in availability of patches ([the most common version of a dependency is probably the most secure](#)), and less-common versions will have fewer eyes); maintenance (older versions will have less open source support, so you'll be doing it all yourself); and even recruiting (no one wants to work on out-of-date versions that don't have current support or examples).

# Data for this report

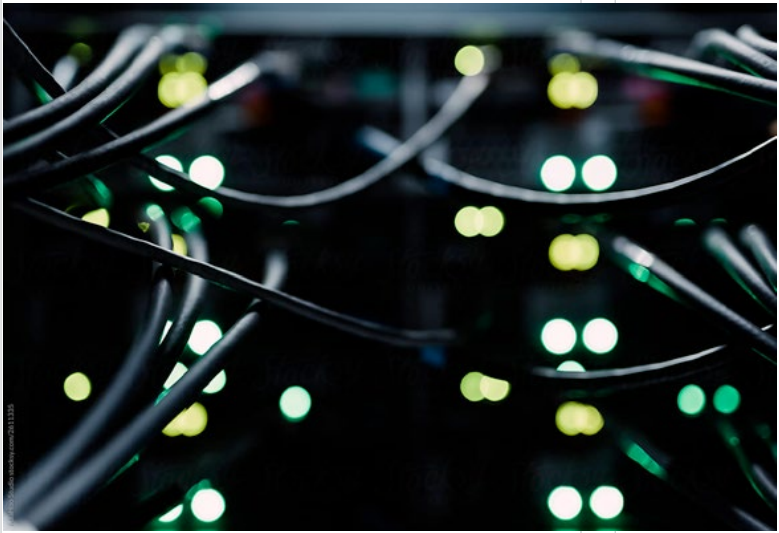

The data for this section comes from GitHub's dependency security features and the six package ecosystems supported. The period of comparison is October 1, 2019 to September 30, 2020 vs. October 1, 2018 to September 30, 2019.

The analysis in this section of the report is based on over 45,000 repositories that meet the following criteria:

- Use one of the six supported package ecosystems

- Are active repositories, which is defined as having at least one contribution in each month from October 2018 through September 2020. This means repositories are only included in the analysis if they were active across two years, thereby excluding new repositories.

- Have dependency graph enabled, which is predominantly public repositories where this is enabled by default

- Are not a fork, classified as spammy, or owned by GitHub staff

//executive summary

The package ecosystems we include in our report and the languages they represent

| Package ecosystem | Language |
|---|---|
| Composer | PHP |
| Maven | Java |
| npm | JavaScript |
| NuGet | .NET |
| PyPI | Python |
| RubyGems | Ruby |

Throughout this section of the report, we reference package ecosystems. A package ecosystem is a collection of libraries packaged in a consistent way in order to make their reuse easy. Most programming languages have a single package ecosystem even if they have more than one package manager.

Our report includes data on the package ecosystems listed, based on [the data we have available](). For example, our analysis does not include data from Java repositories that use the Gradle package manager, or from Python repositories that use Poetry or Conda. While this presents some limitations, we can still gain interesting and meaningful insight into security and best practices.

# Open source security

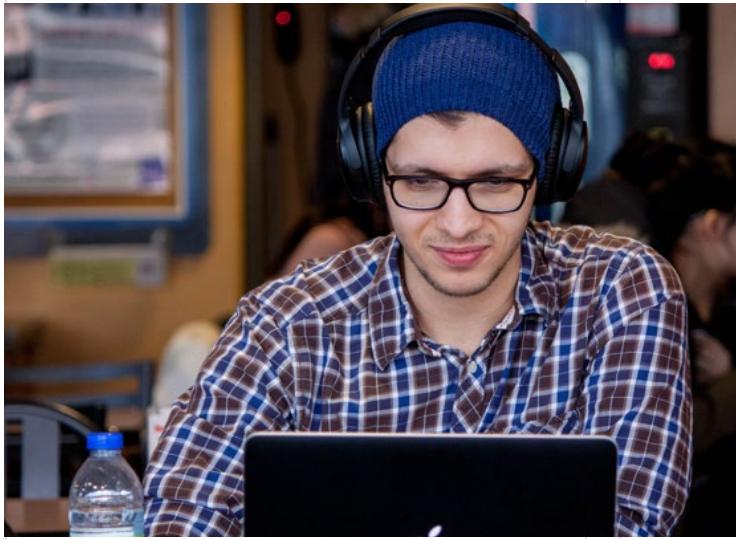

D evelopers worry about introducing security flaws, but that is a risk any time you write code or add a new dependency. The opposite is also a risk: Stale code and outdated dependencies mean attackers have time to methodically attack a system by leveraging every known vulnerability. Malicious attacks exploit flaws in code, and as a result, developers are embracing proactive detection and automation to prevent or limit the impact a bug can have in production. To be successful, we need to consider all vulnerabilities in our code: both the code we write, and the open source software we depend on.

# Surface area

Security vulnerabilities can impact software directly or through its dependencies—any code referenced and bundled to make a software package work. That is, code may be vulnerable either because it contains vulnerabilities, or because it relies on dependencies that contain vulnerabilities. In modern software, 80% or more of most applications' code comes from dependencies, so we looked at package ecosystems and their typical dependency characteristics.

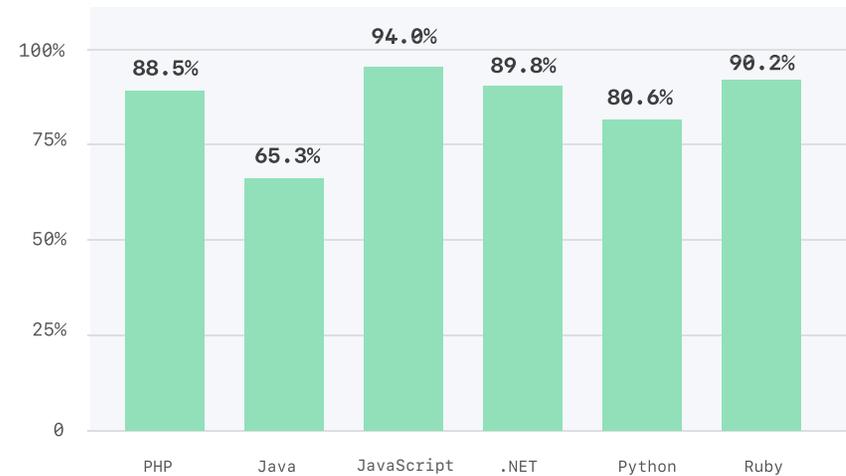

**Percent of active public repositories that use open source software**

- PHP: 88.5%
- Java: 65.3%
- JavaScript: 94.0%
- .NET: 89.8%
- Python: 80.6%
- Ruby: 90.2%

> In its simplest form, a vulnerability is any weakness that can be exploited by an attacker, and can include internal controls, security procedures, implementation, and flaws in computer systems. For this analysis, we focus on vulnerabilities that can be exploited through software.

First, we report the percentage of repositories that reference at least one open source dependency. We see the most frequent use of open source in JavaScript (94%), Ruby (90%), and .NET (90%). We note that Java is likely lower in our dataset because dependency information from repositories using Gradle as a package manager is not available to us. This is about what we would expect, given the way these programs are written and bundled.

Within each repository, we examine the number of dependencies[1] for each package ecosystem. When examining direct dependencies, we find that across all repositories, JavaScript has the highest number of median dependencies (10), followed by Ruby and PHP (nine), and Java (eight), with .NET and Python having the least (six). This shows some variability in median direct dependencies across languages, but not much.

But direct dependencies aren't the whole story. Each direct dependency can itself have dependencies, which may in turn have further dependencies, and so on. We refer to any dependencies that are not "direct" as "transitive dependencies." For languages that include details of their transitive dependencies in lockfiles, and of the repositories with lockfiles,[2] JavaScript has the highest number of median dependencies at 683,[3] followed by PHP (70), Ruby (68), and Python (19).

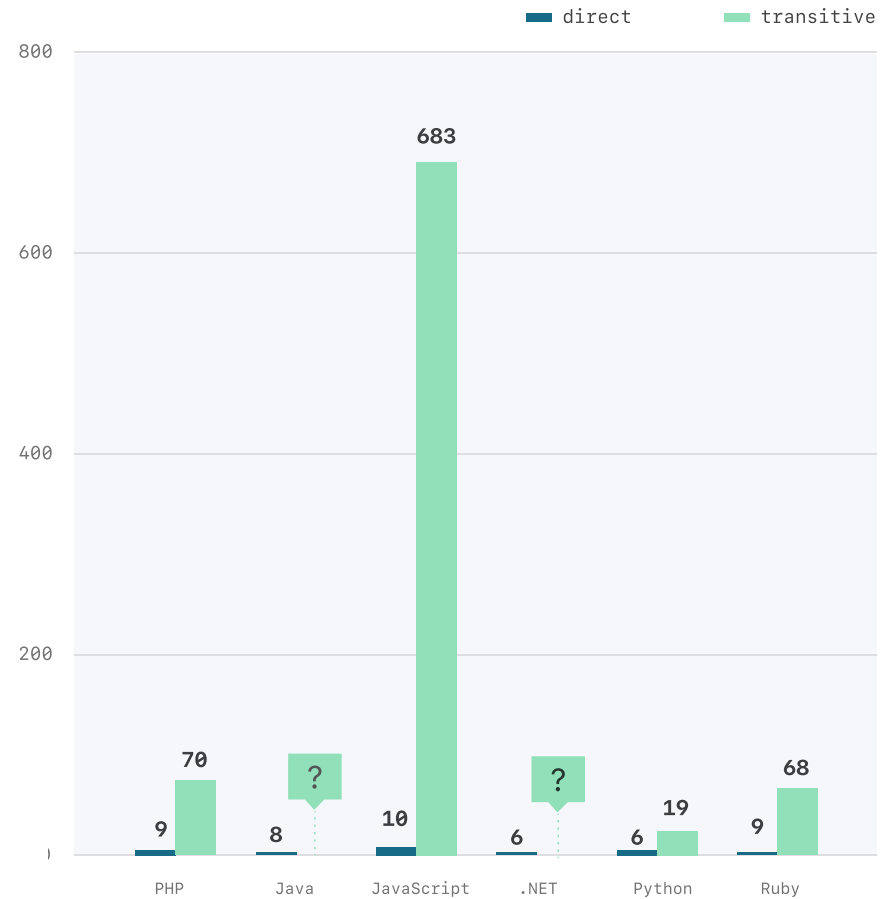

**Median direct and transitive dependencies per repository by package ecosystem**

Note that we have no data on transitive dependencies for Java and .NET (as noted by the "?" in the graphic), and the transitive dependencies bar represents the median exclusively for repositories containing a lockfile.

---

[1] We proxied for direct dependencies by looking for manifest files for that ecosystem (Gemfile, packages.json, pom.xml, etc.) that are not lockfiles. We proxied for direct and transitive dependencies by looking for a lockfile (matching the form ".lock" or "-lock.json").

[2] This includes data for JavaScript (npm), Ruby (RubyGems), PHP (Composer) and Python (for repositories using Pipenv), but not data for the Java (Maven) or .NET (NuGet) ecosystems.

[3] The order of magnitude difference in dependencies between JavaScript and other languages is likely driven by npm's philosophy of "micropackaging" (packaging even one-liner functions as dependencies) together with the small size of the JavaScript standard library and the complex environment in which JavaScript is often used (the web browser). Micropackages are rarely used in applications (i.e., as direct dependencies) but commonly used in libraries, so they show up as transitive dependencies.

# Too malicious to be mistakes: bugdoors and backdoors

Backdoors are software vulnerabilities that are intentionally planted in software to facilitate exploitation. Bugdoors are a specific type of backdoor that disguise themselves as conveniently exploitable yet hard-to-spot bugs, as opposed to introducing explicitly malicious behavior.

The ambiguous nature of backdoors makes them tricky to qualify and establishing intent can be especially challenging. A good backdoor can be indistinguishable from a normal programming mistake. As such, we need to rely on additional indicators to determine the intent of a suspected backdoor event.

The most blatant indicator of a backdoor is an attacker gaining commit access to a package's source code repository, usually via an account hijack, such as 2018's ESLint attack, which used a compromised package to steal a user's credentials for the npm package registry. The last line of defense against these backdoor attempts is careful peer review in the development pipeline, especially of changes from new committers. Many mature projects have this careful peer review in place. Attackers are aware of that, so they often attempt to subvert the software outside of version control at its distribution points or by tricking people into grabbing malicious versions of the code through, for example, typosquatting a package name.

Analysis on a random sample of 521 advisories from across our six ecosystems finds that **17% of the advisories are related to explicitly malicious behavior** such as backdoor attempts. Of those 17%, the vast majority come from the npm ecosystem. While 17% of malicious attacks will steal the spotlight in security circles, vulnerabilities introduced by mistake can be just as disruptive and are much more likely to impact popular projects. Out of all the alerts GitHub sent developers notifying them of vulnerabilities in their dependencies, **only 0.2% were related to explicitly malicious activity**. That is, most vulnerabilities were simply those caused by mistakes.

A big part of the challenge of maintaining trust in open source is assuring downstream consumers of code integrity and continuity in an ecosystem where volunteer commit access is the norm. This requires better understanding of a project's contribution graph, consistent peer review, commit and release signing, and enforced account security through multi-factor authentication (MFA).

# New advisories

Vulnerabilities are reported through advisories, which are available in public databases. This helps developers and open source maintainers secure their software by providing information about issues, fixes, patches, and updates in a centralized location.

## Advisories by package ecosystem

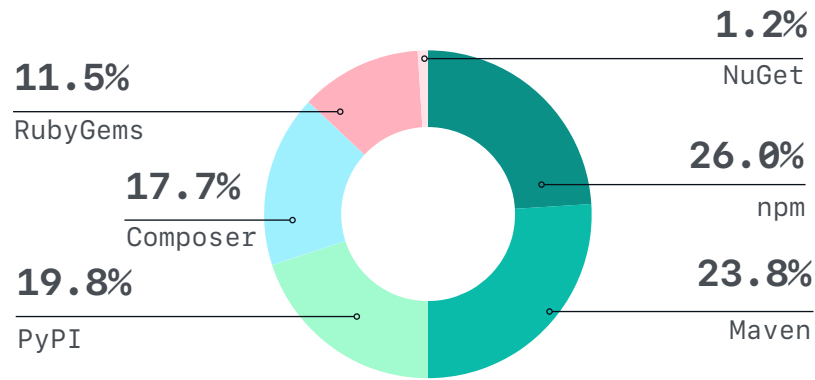

- 1.2% NuGet
- 26.0% npm
- 23.8% Maven
- 19.8% PyPI
- 17.7% Composer
- 11.5% RubyGems

Here we see that npm and Maven have the highest percentage of advisories in the GitHub Advisory Database,[4] with 26% and 23.8%, respectively, and NuGet has the fewest (1.2%). But not all advisories are created equal, as seen when we look at severity.

---

[4] Our initial analysis included a large import of npm advisories to the Advisory Database, when the npm security database was merged into the GitHub Advisory Database following GitHub's acquisition of npm. This import included 738 advisories, and accounted for 24% of the advisories in the database, dominating many trends we were exploring. We exclude this large import from our analysis to ensure the reporting speaks to trends observed generally, but note when npm exhibits different patterns in our analysis. The distribution of advisories by package ecosystem is shown without the large npm import.

# Additional data: security advisories

At this stage of the analysis, we include an additional source of data: GitHub Advisory Database, which contains a curated list of security vulnerabilities that have been mapped to packages tracked by the GitHub dependency graph.

The advisories in this report come from two sources: external ecosystems, which make up 54% of the advisories in our analysis, and maintainer-reported GitHub Security Advisories, which since their introduction in May 2019 already make up the remaining 46%.

External ecosystems include the National Vulnerability Database, RubySec, FriendsOfPHP, and a few other sources that are used occasionally. GitHub carefully verifies third-party feed advisories as well as any maintainer-published advisories for inclusion in the Advisory Database. We evaluate severity, confirm affected version ranges, and check any remediation recommendations.

GitHub Security Advisories allow maintainers to describe, fix, and announce vulnerabilities in their code directly on GitHub. GitHub reviews all published Security Advisories and, whenever appropriate, issues Common Vulnerabilities and Exposures (CVE) IDs for those vulnerabilities. This causes them to be published to the National Vulnerability Database and thus widely available to the global software community. We can do this because we are a CVE Numbering Authority, or CNA.

### Advisories by package ecosystem and severity

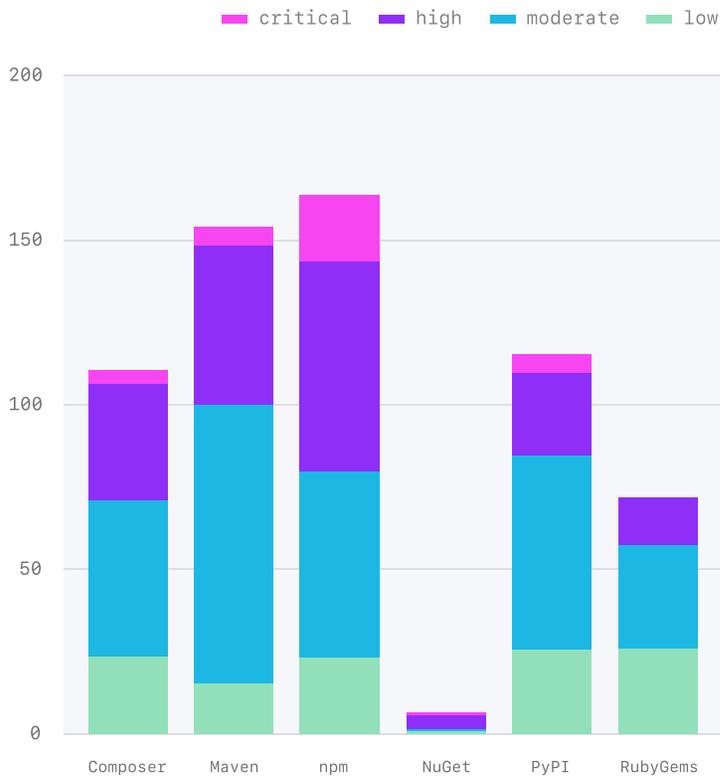

Here we see that npm has the most critical (n=23) and high (n=66) advisories, and that Maven has the most moderate advisories (n=86). RubyGems has no critical advisories and overall, NuGet has the fewest advisories.[5]

## How vulnerabilities are scored

Except for the relatively small proportion of vulnerabilities that are known to be actively exploited in the wild, "severity" is a somewhat subjective concept. Usually, a severity score is assigned by a security expert who looks at the available information and makes a judgment call.

A commonly used tool to help standardize is the Common Vulnerability Scoring System (CVSS). An online CVSS calculator is available from the National Vulnerability Database, (NVD). Four levels are defined in the Common Vulnerability Scoring System (CVSS 3.1): Low, Moderate, High, and Critical. A vulnerability's level depends on factors such as how difficult it is to exploit and how large the impact of a successful exploit could be.

It is relatively easy to assign an accurate score when the vulnerability is in a widely used application and a working proof-of-concept exploit (PoC) is available. But the scoring becomes much more subjective when there is no PoC available, or when the vulnerability is in a library, which means that the exploitability depends on how the library is used. Like all CVE Numbering Authorities, GitHub strives to objectively follow the CVSS when assigning a severity. We also work proactively with NVD on the CVMAP process, which uses US government security researchers to objectively evaluate severity scores across all vulnerabilities submitted to the NVD.

---

[5] Whereas other ecosystems have detailed, community-curated ecosystem sources for their advisories, these are still fairly limited for NuGet and critically, are not machine readable. So although NuGet seems to have fewer advisories than other ecosystems, that doesn't necessarily mean it's safer.

# Security alerts

An important part of the vulnerability remediation process is understanding and tracking software inventory, matching against security advisories, and then alerting when relevant vulnerabilities appear. This involves identifying the vulnerable component and the corresponding vulnerability so teams can take the appropriate action to ensure their code is secure.

### Additional data: alerting

We now add an additional source of data to our analysis: Dependabot. Dependabot alerts developers to vulnerable dependencies in public repositories by default, and developers may opt out. In contrast, private repositories are opt-in, and developers must enable Dependabot alerts at the individual or organization level. Because of this, not all repositories get alerts. For this analysis, we capture alerts that were sent to developers.

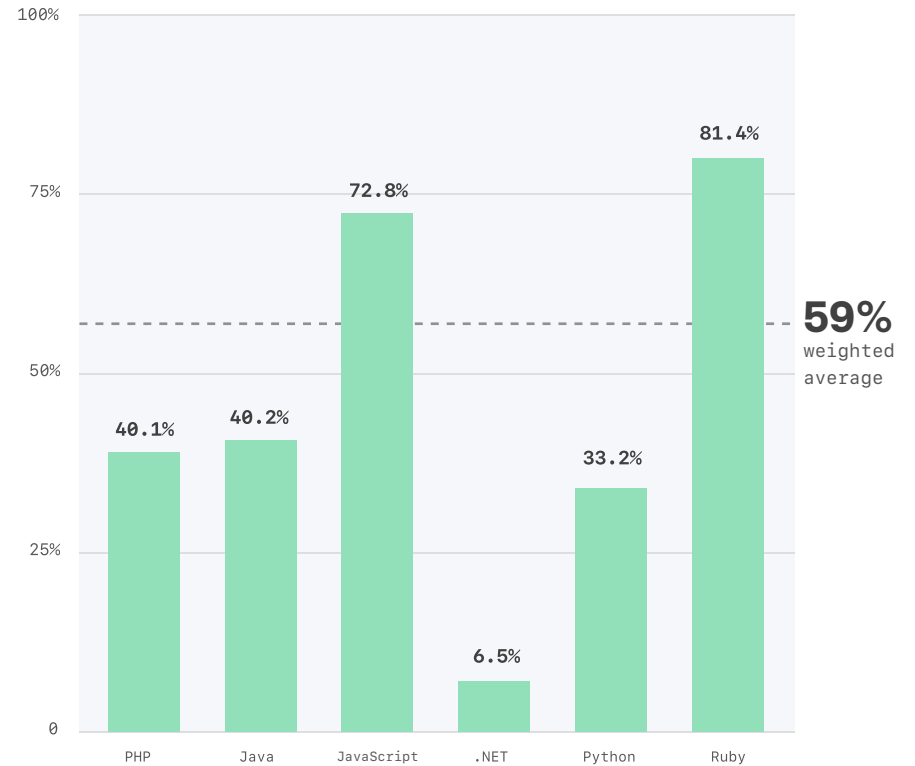

**Percentage of active repositories that received Dependabot alerts**

- PHP: 40.1%
- Java: 40.2%
- JavaScript: 72.8%
- .NET: 6.5%
- Python: 33.2%
- Ruby: 81.4%

**59%** weighted average

Overall, active repositories with a supported package ecosystem have a **59% chance of getting a security alert**. Broken down by package ecosystem, we see the repositories most likely to get an alert use RubyGems (81%) and npm (73%).

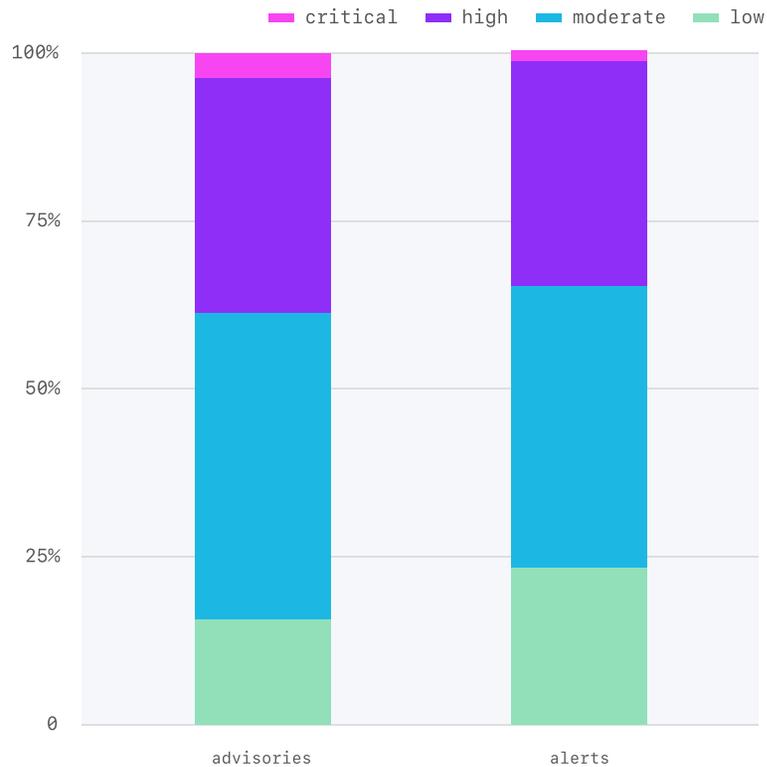

**Advisories by severity level and automated alerts by severity level** (via Dependabot)

Here we see a breakdown of advisories by severity level and automated alerts by severity level. Note they differ somewhat from the proportions of advisories we see in the Advisory Database. The biggest differences lie in low severity (24% of alerts, which is higher than the proportion of advisories in the Advisory Database) and critical severity (two percent of alerts, which is a lower proportion than the advisories in the Advisory Database). This means that users are receiving a disproportionate amount of alerts for lower-severity vulnerabilities, although this could mean that without sufficient differentiation, the rare, more-severe alerts drown out in the noise. It also means that the most critical vulnerabilities aren't occurring in as widespread components, and so affect fewer users from the get-go.

To see how security alerts are distributed across package ecosystems, and how they differ from the advisories available in the Advisory Database, we show both distributions.

## Alerts per package ecosystem

We break down alerts per package ecosystem for the previous year, and see that the severity of a vulnerability found in a dependency is not very correlated with how many people use that dependency.

Over the last 12 months, the vulnerabilities found in the most-used npm packages were low or moderate severity. It's tempting to conclude that severity is negatively correlated with popularity, perhaps lending credibility to the theory that many eyes make all bugs shallow—meaning the most critical vulnerabilities are caught in code review. However, a quick look at the severity distribution for Composer alerts is enough to dissuade us of that misconception; It had critical and high-severity vulnerabilities in its biggest packages. The truth appears to be that it is just as easy for a critical-severity vulnerability to make it through code review as a low-severity one.

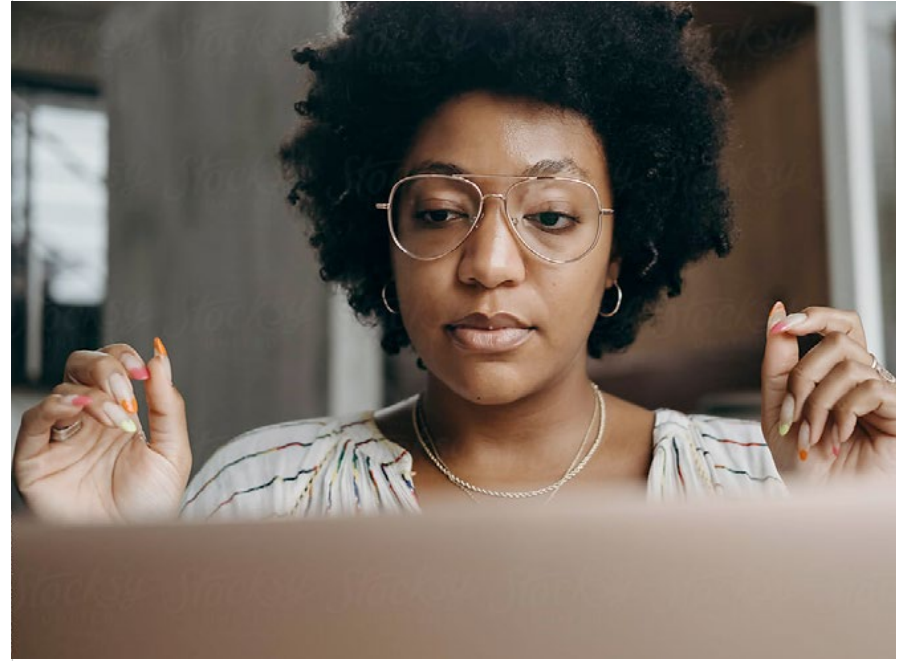

# Lifecycle of open source vulnerabilities

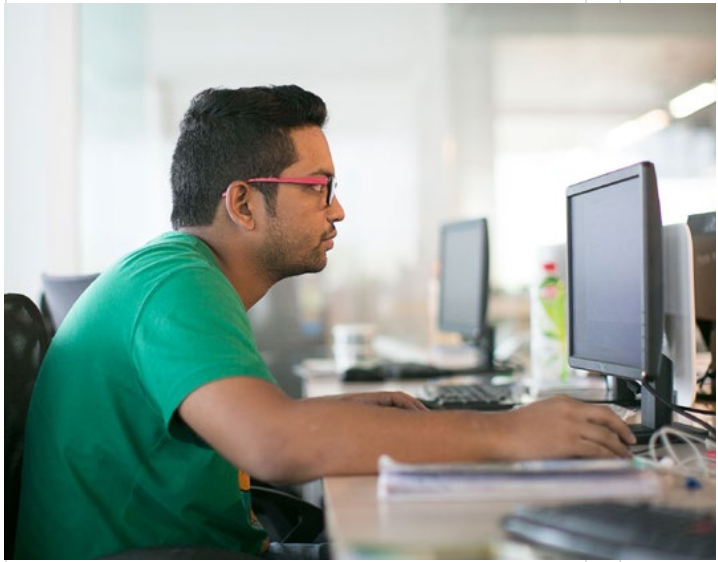

S ecurity vulnerabilities are an important part of software development and delivery, and application security professionals help teams and organizations secure their code and systems. In this section, we look at the lifecycle of a vulnerability and show how best practices can help remediate vulnerabilities faster, resulting in more secure and reliable software.

The four steps to open source vulnerability remediation are:

1. A vulnerability is identified and reported.
2. The maintainer fixes the vulnerability and releases a new version.
3. Security tooling alerts end users of a security update.
4. Developers update to the fixed version.

Anyone in the open source community can identify a vulnerability (step one), though it is usually security researchers. Maintainers then take the lead on creating a fix and releasing a security update (step two). End users of the dependency are notified (step three) after the maintainer or a security researcher requests a CVE for the vulnerability and security tools add it to their databases. Those end users then update their code to use the newly released fixed version (step four).

### The full lifecycle of a vulnerability

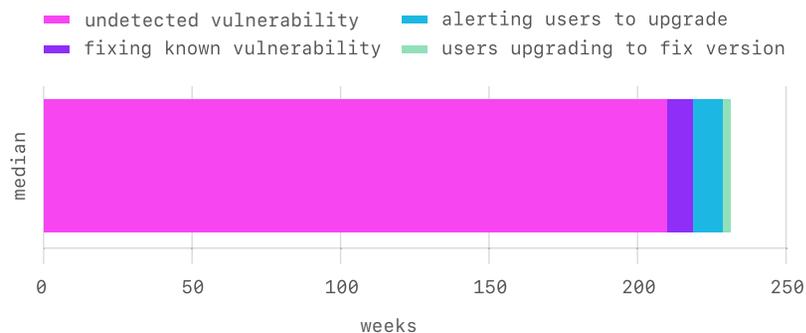

A vulnerability typically goes undetected for 218 weeks (just over four years[6]) before being disclosed.[7] From there, it typically takes 4.4 weeks for the community to identify and release a fix for the vulnerability, and then 10 weeks to alert on the availability of a security update.[8] We find that for repositories that do apply the fix, it typically takes one week to resolve.[9] There is an opportunity to shorten the life of a vulnerability by focusing efforts on time to detect. This highlights two things: the importance of focusing efforts on time to detect, and that there are likely a large number of undiscovered vulnerabilities in our open source software today. If our development efforts introduce them at a constant rate, the rate of discovery significantly lags behind the rate of introduction.

There are differences by package ecosystem, advisory database, and which method a security team uses for alerts and remediations. We'll investigate each stage in more detail, with our analysis focusing on RubyGems and npm because of the ample data available.

---

[6] Four years may seem like a long time before a vulnerability is detected, but it's not unheard of. While different from our own analysis of all vulnerabilities, RAND reports that zero-day vulnerabilities—those that are unknown to anyone but hackers who can exploit them—typically go undetected for five years.

[7] The method used to proxy the timeline for a vulnerability to be discovered likely skews long. Because fixes are often applied to code "at or before version X," we captured the timeline for all of those potentially affected versions. While the vulnerability is often introduced in a commit much closer to the fixed version, it's infeasible to identify without root cause analysis, and not at scale for the purposes of this report.

[8] These ten weeks to alert are the result of many factors, including times for import and curation across several communities.

[9] For this analysis, we focused on the first 25% of active repositories to patch their software; this represents a typical timeline for repositories that intend to update.

## The worst vulnerabilities of 2020

Which vulnerability is the worst as of November 2020? It depends on how you define "worst." Some obvious candidates are CVE-2020-0601 (aka Curveball), CVE-2020-0796 (aka SMBGhost), and CVE-2020-1472 (aka Zerologon). These vulnerabilities were severe in terms of the number of developers they affected and their potential impact on vulnerable networks and endpoints. These could be the worst because they are severe vulnerabilities that require urgent attention from systems administrators.

But another definition of worst is the vulnerability that has the most impact on project maintainers. By this definition, a strong candidate for most-impactful bug of the year is CVE-2020-8203 (Prototype Pollution in lodash). That vulnerability is single-handedly responsible for over five million Dependabot alerts. That's because lodash is one of the most widely used npm packages. Furthermore, Prototype pollution is a potentially severe vulnerability, which in the worst case could lead to remote code execution where the zipObjectDeep method is used. Developers are strongly advised to upgrade to the latest version of lodash.

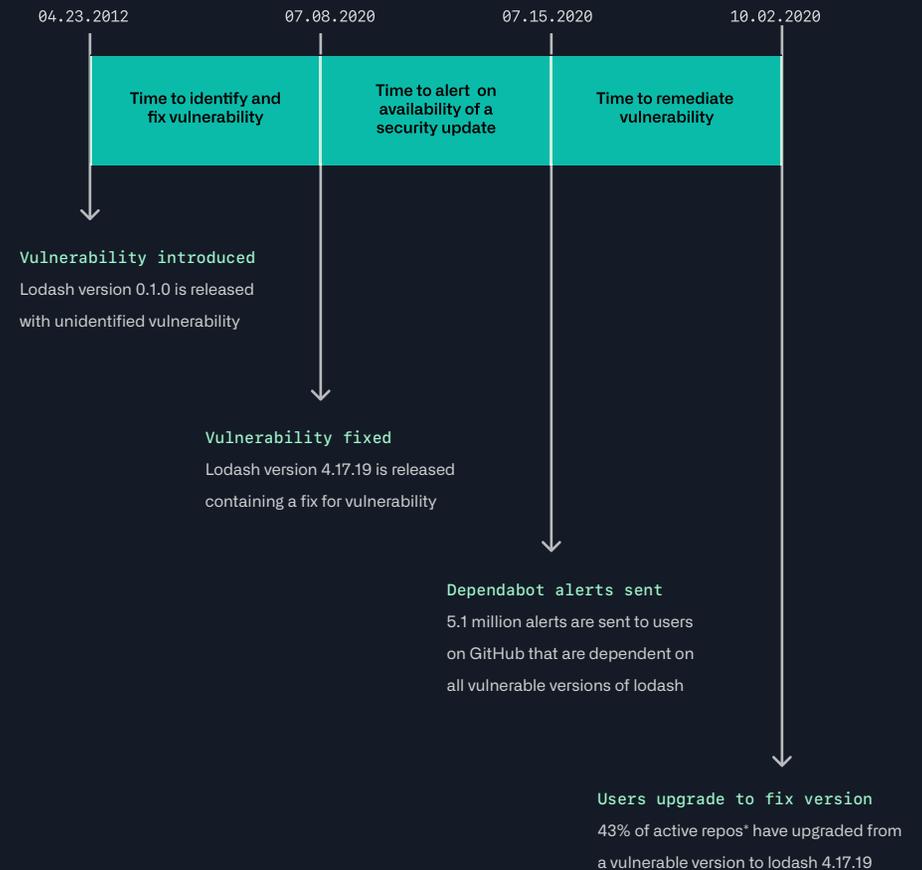

Timeline for lodash vulnerability

04.23.2012 — 07.08.2020 — 07.15.2020 — 10.02.2020

Time to identify and fix vulnerability | Time to alert on availability of a security update | Time to remediate vulnerability

**Vulnerability introduced**
Lodash version 0.1.0 is released with unidentified vulnerability

**Vulnerability fixed**
Lodash version 4.17.19 is released containing a fix for vulnerability

**Dependabot alerts sent**
5.1 million alerts are sent to users on GitHub that are dependent on all vulnerable versions of lodash

**Users upgrade to fix version**
43% of active repos* have upgraded from a vulnerable version to lodash 4.17.19

* An active repo is defined as one with a push in the week before the Dependabot alerts were sent out

**FOR SECURITY RESEARCHERS AND MAINTAINERS**

# Identify and fix a vulnerability

The time between a vulnerability being introduced into an ecosystem and when security researchers and maintainers identify a fix is typically seven years for RubyGems and five years for npm. This is because software vulnerabilities often go unnoticed and undetected. In addition, many teams may lack the expertise—or simply the time—to find vulnerabilities in their code, focusing on developing core functionality instead.

Looking at severity, we find that critical vulnerabilities are disclosed faster. This is obviously good news, but it's not immediately clear what drives this faster timeline, and is worth more research.

### Time to identify and fix a vulnerability, distribution in years; npm and RubyGems

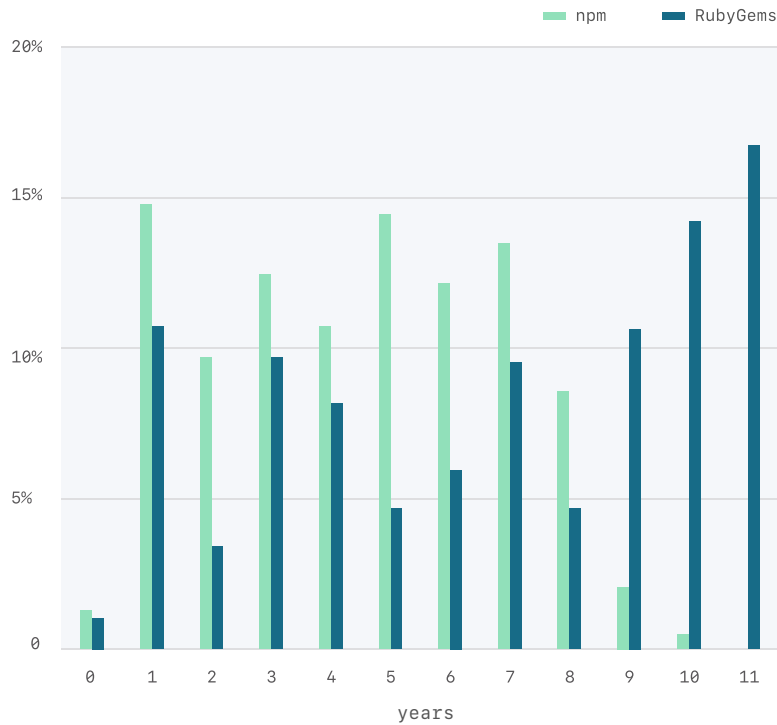

### Time to identify and fix a vulnerability, by severity

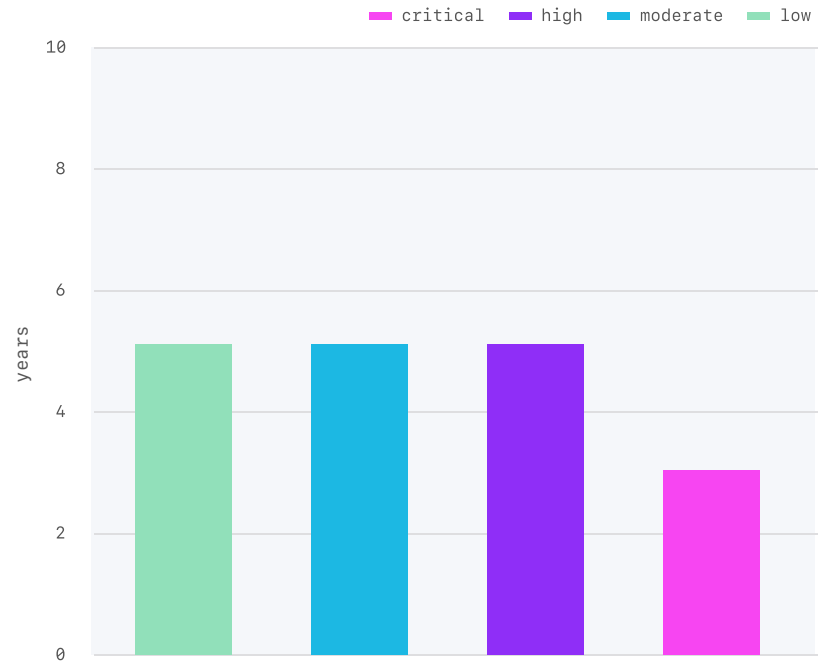

FOR MAINTAINERS ON GITHUB

# Alert on availability of security update

Time to alert is an important component for any security professional, and beyond that, for their broader DevOps teams. Other than finding the vulnerability themselves, receiving an alert about a vulnerable dependency is the first opportunity that teams have to respond. Once a fix is discovered, the team can patch any vulnerable code and upgrade impacted systems.

> The alerting mechanism and data we used for this analysis comes from the GitHub Advisory Database and Dependabot alerts.

Time to alert differs based on where the advisory originates: imported from external sources or submitted directly to the GitHub Advisory Database. The difference stems from the process used to submit advisories. Because the GitHub Advisory Database receives submissions directly, maintainers can draft an advisory even before the fix is ready, and obtain a CVE directly from GitHub. Once published, this can send out alerts much faster, typically within one week. In contrast, imported advisories must go through other channels before finally making it into the central repository, which can introduce delays. These other channels may require passing through several intermediaries in sequence, obtaining a CVE, publishing a fix, and submitting to the NVD. The distribution of alerts shows that repository advisories are strongly skewed toward fast alerting, while imported advisories are typically alerted after 20 weeks and have a longer tail.

### Weeks from fix version to Dependabot alerts sent

Cumulative distribution in weeks, time from vulnerability becoming known (proxied by a fixed version being released) to inclusion in the GitHub Advisory Database

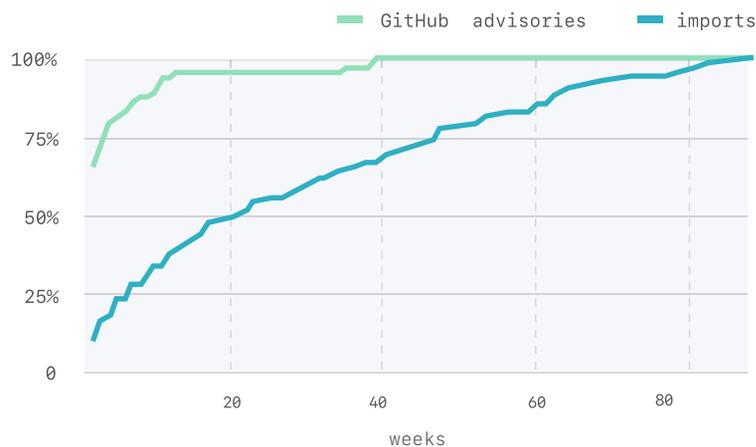

# Remediate the security update

If software was a fairytale, teams would get an alert, install the patch, and their systems would be secure. But software is complex, and knowing a patch is available is often just a beginning to remediation. Patching early and often is the best security strategy, but isn't always an option. Teams may need to wait to ensure they don't interrupt operations, may have too many patches to merge at once, or may have to work around features or legacy platforms that won't yet support the patch.

Knowing that patching security vulnerabilities is not straightforward, we investigated the time to remediate. This analysis includes time to resolve an alert for any repository that did resolve.

**Cumulative percent of npm and RubyGems Dependabot alerts resolved by severity over time**

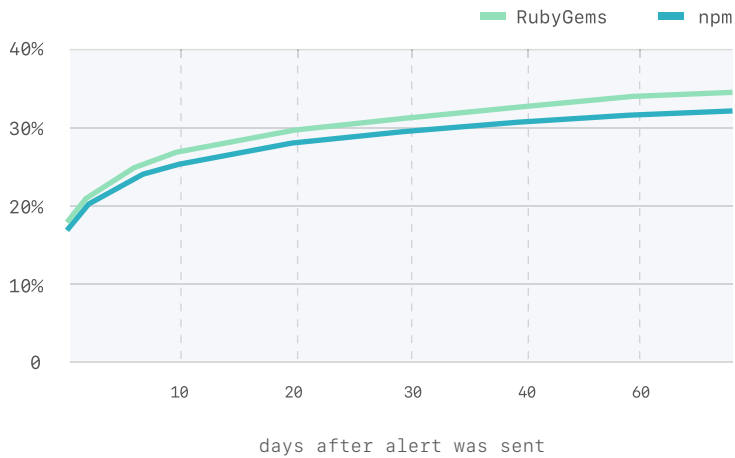

days after alert was sent

Across all repositories, both RubyGems and npm alerts see resolution rates close to 20% resolved within a day. This rate steadily increases over time, reaching about 30% resolved within a month after alert.

**Cumulative percent of Dependabot alerts resolved by severity over time**

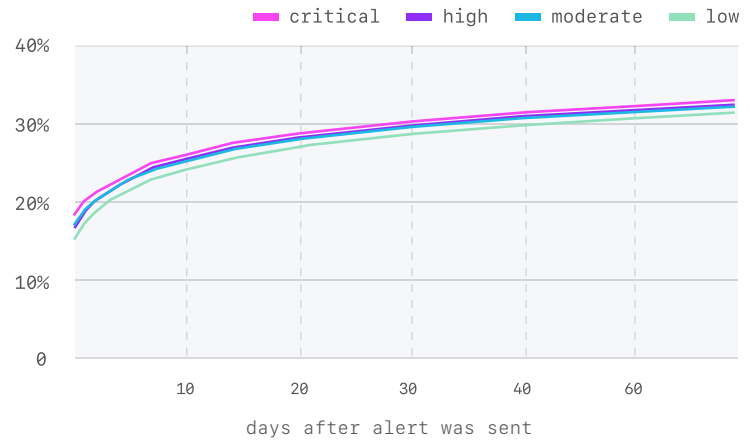

days after alert was sent

Developers react faster to more severe issues, but the gap isn't huge—alerts of all severities see resolution rates close to 20% resolved within a day. This rate steadily increases over time, reaching about 30% resolved within a month after alert.

# More secure with automation

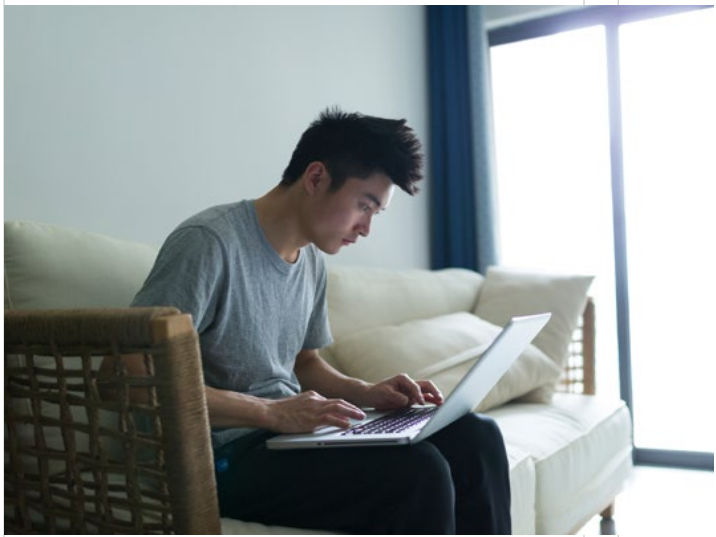

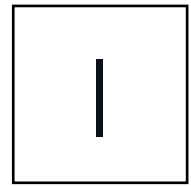

Is software getting more secure? Are we getting better at alerting and are teams getting better at resolving found vulnerabilities? These are difficult to answer, in part because software is growing and changing.

Software and systems are evolving as we build new features and maintain our infrastructure, which means our attack surfaces are evolving. This pushes attackers to stay active and ready, because a surface they have previously exploited could be replaced with a new feature or patched at any time. At the same time, new people are joining our teams and projects and learning how to secure software and systems.

Automation, such as automatic dependency version updates with Dependabot, provides another opportunity for developers to secure their code. By automating security practices, developers share their expertise with their community, remove security and engineering silos, and scale their expertise. It allows developers to explore opportunities while providing critical infosec continuity. And teams can leverage the power of the broader infosec community to identify and remediate security vulnerabilities in their codebases.

Many developers use open source to create and build projects faster, and research by DORA finds that elite performers are 1.75 times more likely than low performers to make extensive use of open source software.

While some worry that open source code may have unseen dependencies and vulnerabilities, security is always a concern when working with software. Our analysis shows that potential vulnerabilities found[11] scale with the number of lines of code written. The power and promise of open source is in the power of the community. By joining forces with millions of developers to not only build software packages but also identify and fix vulnerabilities, we can build software more quickly and more securely. The key is to leverage automated alerting and patching tools to secure your software quickly.

**Potential vulnerabilities found in source code scale with lines of code written**

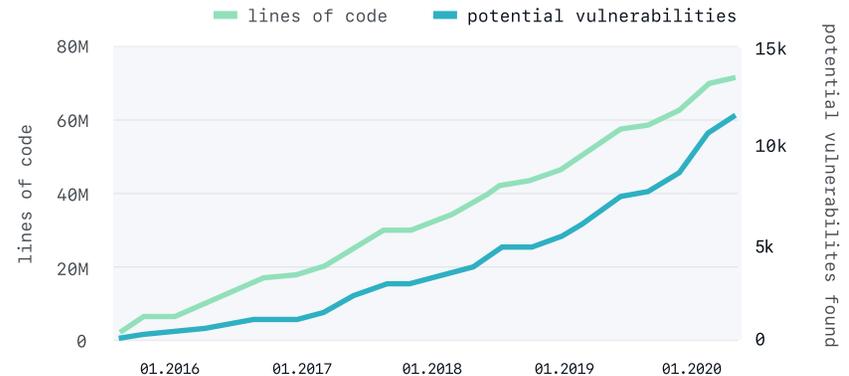

Are we introducing fewer vulnerabilities in the code we write today than we did in the code we wrote in the past? An analysis of commits to open source repositories suggests not.

By running static analysis on historic commits to a project, we can see when new potential vulnerabilities were introduced. We ran CodeQL, GitHub's static analysis security tool, on each commit to several thousand popular open source projects over a five year period to see if the rate at which vulnerabilities are introduced has changed over time. The result is the graph above, which suggests a line of code written in 2020 is just as likely to introduce a security vulnerability as one written in 2016.

---

[11] These potential security vulnerabilities are static analysis alerts. The analysis applies the same static analysis engine and queries to every commit to see how the number of alerts changed over time. This represents a relatively unbiased proxy for vulnerabilities.

# Automating vulnerability remediation: shifting left

DevSecOps professionals proclaim "shifting left" is a superpower, saying that building security into the development process amplifies the expertise of infosec professionals. But how can these teams shift left and build in security?

Research from DORA points to automation that makes it easy for teams to integrate security into the development process as a predictor for high performance. Our own analysis found that repositories that automatically generated a pull request to update to the fixed version patched their software in 33 days, which is **13 days faster than those who did not, or 1.4 times faster**. Using automation is an important best practice: Teams who automate both the pull request and have in place extensive continuous integration checks for security patches report these are critical to fast updates.

Sonatype also found that high-performing software development teams are 4.9 times more likely to successfully update dependencies and fix vulnerabilities without breakage.

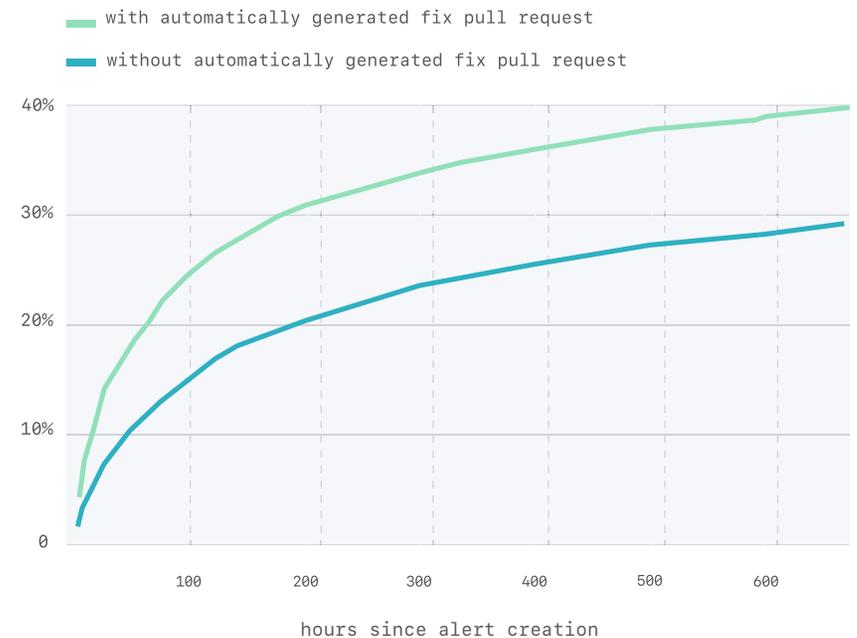

**Percent of Dependabot alerts resolved by hour**

Software security is everyone's job. And the effort is worth it: having good automation and patching practices makes it easier and safer to integrate fixes into our development work.

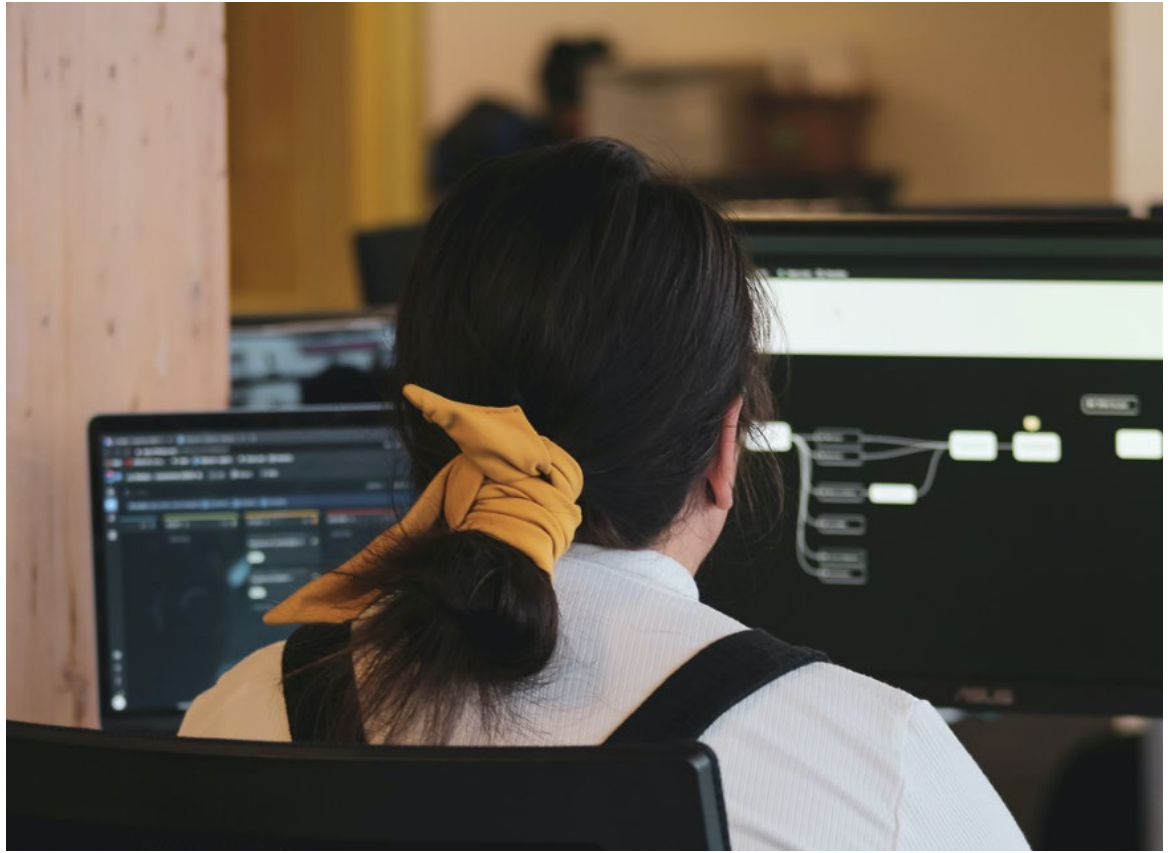

**For more insights about how we work**

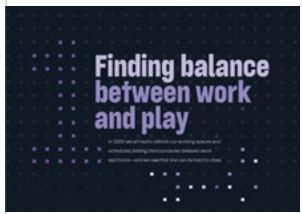

Finding balance

**Productivity report →**

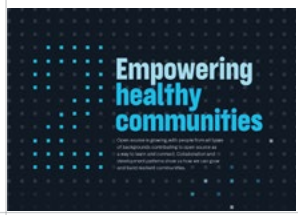

Empowering communities

**Community report →**

# The security of open source is mission-critical

# Glossary

The methodology and data used for analysis is described throughout the report

**Active repository**
An active repository is one that has at least one contribution in each month during the time period of analysis.

**Dependency graph**
This feature lists all dependencies for a repository and helps identify known vulnerabilities.

**Developers**
Developers are individual accounts on GitHub, regardless of their activity.

**GitHub Advisory Database**
An advisory database contains all curated CVEs and security advisories that have been mapped to a package tracked by the GitHub dependency graph.

**Location**
Country information for developers is based on their last location, where known. For organizations, we take the best-known location information either from the organization profile or the most common country organization members are active in. We only use location information in aggregate form to look at things like trends in growth in a particular country or region. We don't look at location information granularity finer than country level.

**Open source projects**
Open source projects are public repositories with an open source license.

**Organizations**
Organization accounts represent collections of people on GitHub. These can be paid or free, big or small, businesses or nonprofits.

**Projects and repositories**
We use projects and repositories interchangeably, although we understand that sometimes a larger project can span many repositories.

**Vulnerabilities**
This is a problem in a project's code that could be exploited to damage the confidentiality, integrity, or availability of the project or other projects that use its code. Vulnerabilities vary in type, severity, and method of attack.

// acknowledgements

Many thanks to our data scientists, contributors, and reviewers. Each is listed alphabetically by type of contribution.

------------------------

**Authors:** Nicole Forsgren
with contributions from Bas Alberts, Kevin Backhouse, Grey Baker

**Data Scientists**: Bas Alberts, Grey Baker, Greg Cecarelli, Derek Jedamski, Scot Kelly, Clair Sullivan

**Reviewers:** Grey Baker, Dino Dai Zovi, Denae Ford, Maya Kaczorowski, Alex Mullans, Kelly Shortridge

**Copyeditors:** Leah Clark, Cheryl Coupé, Stephanie Willis

**Designers:** Siobhán Doyle, Aja Shamblee

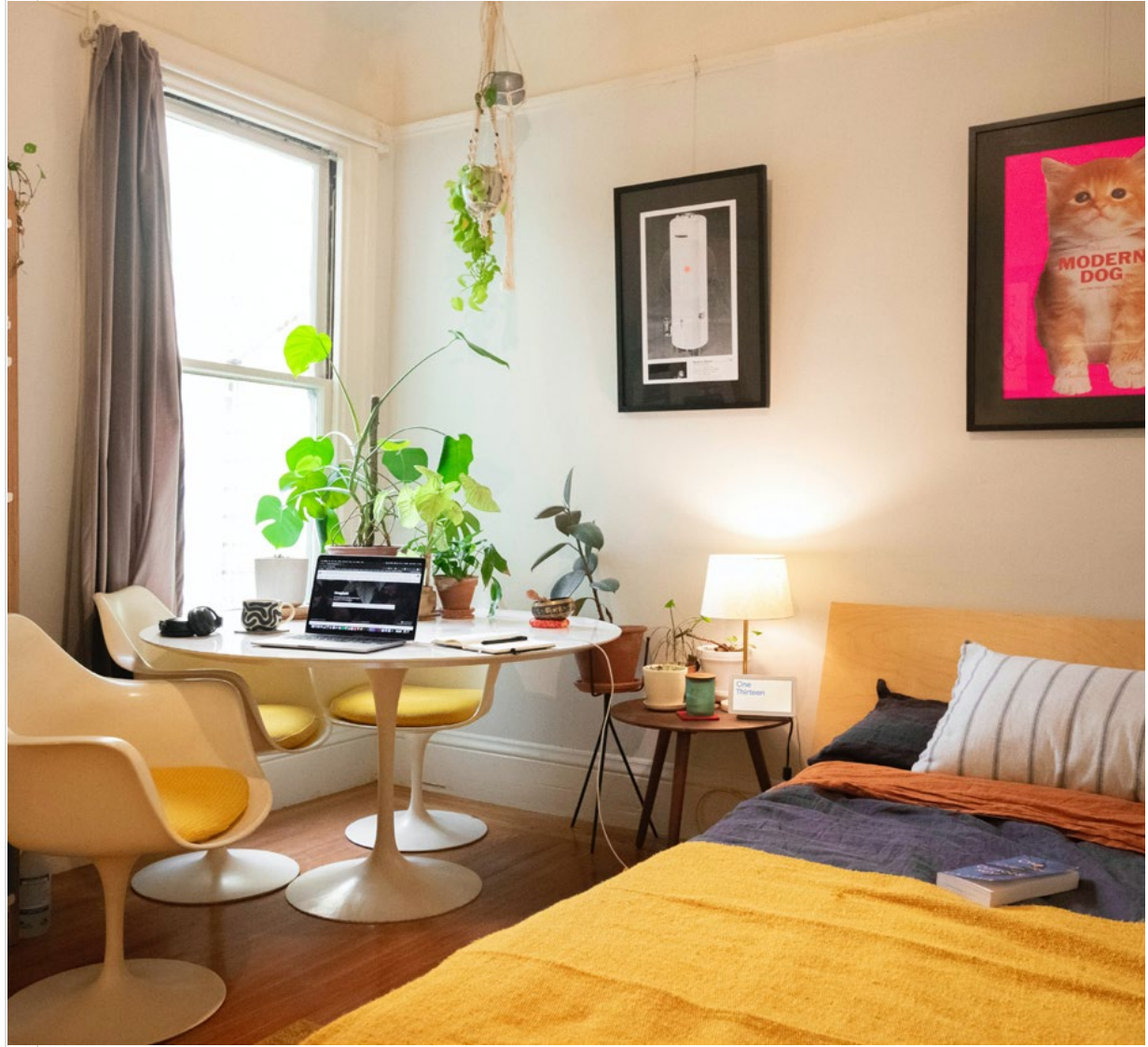

**Alerts sent for each advisory, log scale (via Dependabot)**

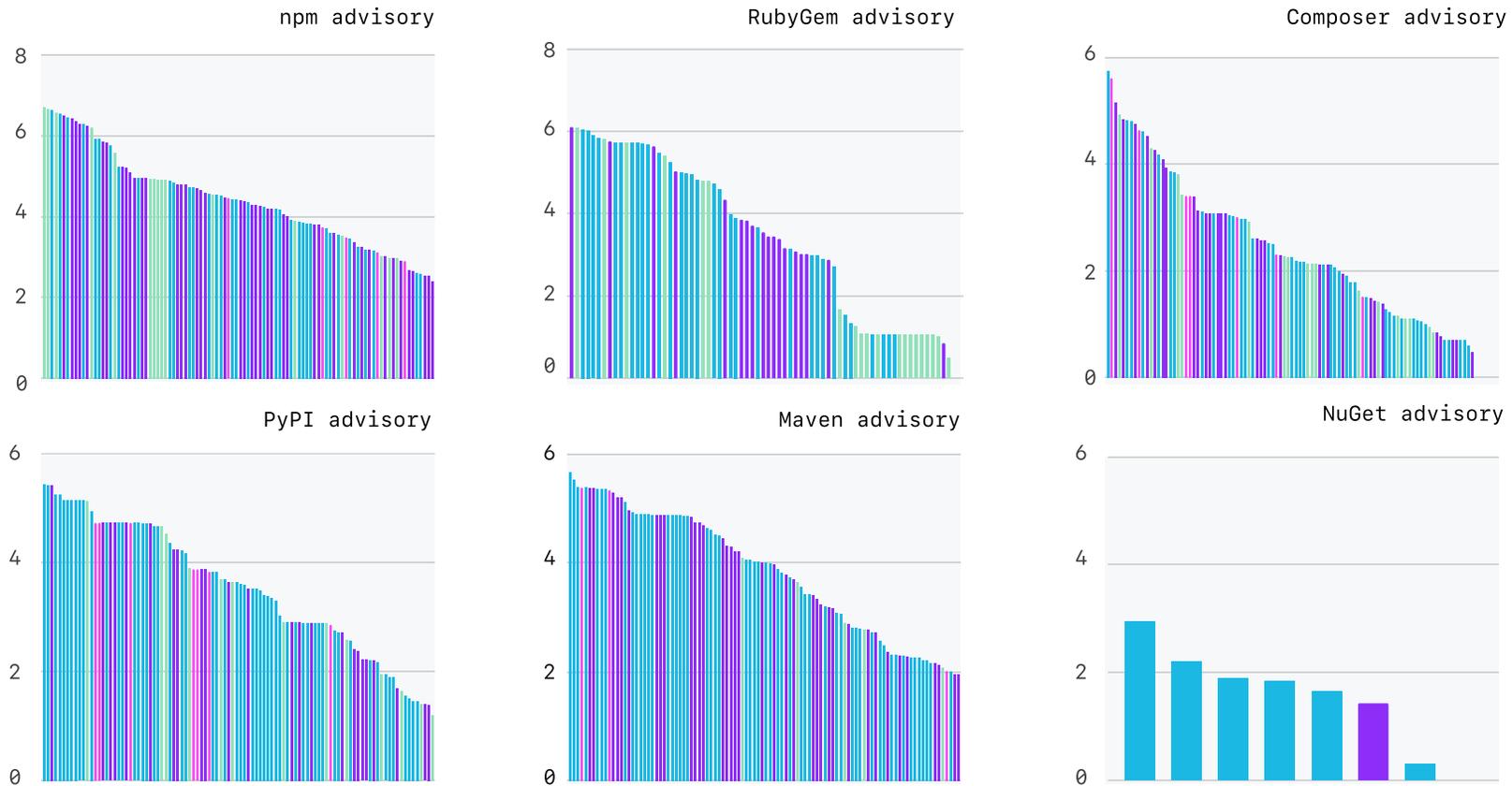

Alerts sent for each language, log scale